\documentclass[a4paper,11pt]{article}
\usepackage{graphicx} % Required for inserting images
\usepackage{jheppub}
\usepackage{amssymb}
\usepackage{amsmath}
\usepackage{soul}
\usepackage{hyperref}	%This is for pdftex
\hypersetup{colorlinks,bookmarksopen,bookmarksnumbered,citecolor=blue,
linkcolor=blue,pdfstartview=FitH,urlcolor=blue}
\usepackage{xcolor}
\usepackage{subcaption}
\usepackage{fontawesome5}

\title{\boldmath High Frequency Gravitational Wave Bounds from Galactic Neutron Stars}

\author[a]{V. Dandoy,}
\author[b%b,1
]{T. Bertólez-Martínez%,\note{Corresponding author.}
,}
\author[c%b,1
]{F. Costa.%,\note{Corresponding author.}
}
\affiliation[a]{Service de Physique Theorique, C.P. 225, Universite Libre de Bruxelles,\\
Boulevard du Triomphe, B-1050 Brussels, Belgium}
\affiliation[b]{Departament de F\'isica Qu\`antica i Astrof\'isica and Institut de Ci\`encies del Cosmos, Universitat de Barcelona, Diagonal 647, E-08028 Barcelona, Spain}
\affiliation[c]{Institute for Theoretical Physics, 
	Georg-August University G\"ottingen,\\
	Friedrich-Hund-Platz 1, G\"ottingen, D-37077 Germany}
\emailAdd{virgile.dandoy@ulb.be}\emailAdd{antoni.bertolez@fqa.ub.edu}
\emailAdd{francesco.costa@theorie.physik.uni-goettingen.de}
\abstract{High-Frequency Gravitational Waves (HFGWs) constitute a unique window on the early Universe as well as exotic astrophysical objects. While the current gravitational wave experiments are more dedicated to the low frequency regime, the graviton conversion into photons in a strong magnetic field constitutes a powerful tool to probe HFGWs. In this paper, we show that neutron stars, due to their extreme magnetic field, are a perfect laboratory to study the conversion of HFGWs into photons. Using realistic models for the galactic neutron star population, we calculate for the first time the expected photon flux induced by the conversion of an isotropic stochastic gravitational wave background in the magnetosphere of the ensemble of neutron stars present in the Milky Way. We compare this photon flux to the observed one from several telescopes and derive upper limits on the stochastic gravitational wave background in the frequency range $10^8 \, \rm Hz$ - $10^{25}\, \rm Hz$. We find our limits to be competitive in the frequency range $10^8 \, \rm Hz$ - $10^{12}\, \rm Hz$. }

\begin{document} 
\maketitle
\flushbottom

\section{Introduction}

The search for low-frequency (nHz-kHz) Gravitational Waves (GWs) has been strongly motivated by theoretical predictions in the last decades. This has naturally led to a series of experimental projects aiming to detect GWs below the $\sim \rm kHz$. Among them, the Laser Interferometer Gravitational-Wave Observatory (LIGO)~\cite{LIGOScientific:2016aoc} first observed GWs in the kHz range from binary black holes. More recently, the Pulsar Timing Array (PTA) community (NANOGrav~\cite{NANOGrav:2023gor}, PPTA~\cite{Reardon:2023gzh}, EPTA~\cite{EPTA:2023fyk}, and IPTA~\cite{Antoniadis:2022pcn}) has claimed strong evidence of a Stochastic Gravitational Wave Background (SGWB) in the nHz regime. The latter is strongly motivated by supermassive black holes binaries~\cite{Middleton:2020asl, Phinney:2001di}. 
%,even though some early Universe sources have been found to explain similarly well the signal. 
Early Universe sources have been shown to offer a similar level of explanation to the SGWB~\cite{Vagnozzi:2020gtf,Chen:2021ncc,Sakharov:2021dim,Benetti:2021uea,Ashoorioon:2022raz,Wu:2023pbt,IPTA:2023ero,Wu:2023dnp,Dandoy:2023jot,Madge:2023dxc,Kawai:2023nqs,Vagnozzi:2023lwo}.\\
In analogy to the electromagnetic spectrum that covers orders of magnitudes in frequency, GWs could be produced over a large frequency range, each of them being associated with different physical phenomena. Interestingly, GWs with frequencies larger than the ones currently searched for in experiments are difficult to motivate with known astrophysical objects or standard cosmology~\cite{Aggarwal:2020olq}. Any detection of such GWs would, therefore, be a hint for exotic early Universe sources; phase transitions, preheating after inflation, oscillons, cosmic strings, or exotic astrophysical objects, for instance, primordial black holes \cite{Ghiglieri:2015nfa, Wang:2019kaf,Ghiglieri:2020mhm, Ghiglieri:2022rfp, Khlebnikov:1997di, Easther:2006gt, Garcia-Bellido:2007nns, Anantua:2008am,Dolgov:2011cq, Dong:2015yjs,Costa:2022lpy,Costa:2022oaa,Gehrman:2023esa} (see Ref. \cite{Caprini:2018mtu} for a review of different sources). These primordial GWs are expected to propagate freely in the Universe to form a SGWB, as in the case of the background potentially observed by the PTA.\\ %even
However, High-Frequency Gravitational Waves (HFGWs) have been significantly less explored than the low-frequency range. The main reason for this relies on the fact that GW detectors are usually sensitive to GWs with wavelengths comparable to their size. Targeting HFGWs, therefore, requires a radically different experimental setup (see Refs.~\cite{Nishizawa:2007tn, Grishchuk:2003un, Braginsky:1978yp, Aggarwal:2020olq, Ito:2019wcb, Ito:2020wxi,Kahn:2023mrj} for a non-exhaustive list of recent proposals).\\
Nonetheless, the detection of HFGWs may be possible via the (inverse) Gertsenshtein effect \cite{Raffelt:1987im, Boccaletti:1970pxw}: similarly to axions, GWs in the presence of a magnetic field can convert into photons. This has naturally led to use the existing axion experiments as HFGW detectors and translating the existing constraints on the axion to photon coupling into constraints on the GWs abundance $\Omega_{\rm gw}$ (usually expressed in terms of the characteristic strain $h_c$)~\cite{Ringwald:2020ist, Ejlli:2019bqj, Berlin:2021txa, Domcke:2022rgu, Tobar:2022pie}, see Fig.~\ref{fig: Constraints}.\\
More recently, the conversion of a high-frequency SGWB into photons in astrophysical environments has been considered. For instance, Refs.~\cite{Dolgov:2012be, Domcke:2020yzq} studied the impact on the CMB/X-ray measurements of the conversion of MHz-GHz GWs in the cosmic magnetic field and extracted strong constraints on the GWs abundance. Similar conversion has been studied in the magnetosphere of solar system planets~\cite{Liu:2023mll} and the Crab/Gemini pulsars~\cite{Ito:2023fcr}, as well as in the geomagnetic field, and the galactic and intergalactic magnetic fields~\cite{Ramazanov:2023nxz, Ito:2023nkq}.\\
\begin{figure}[t!]
    \centering
    \includegraphics[width = 0.45\textwidth]{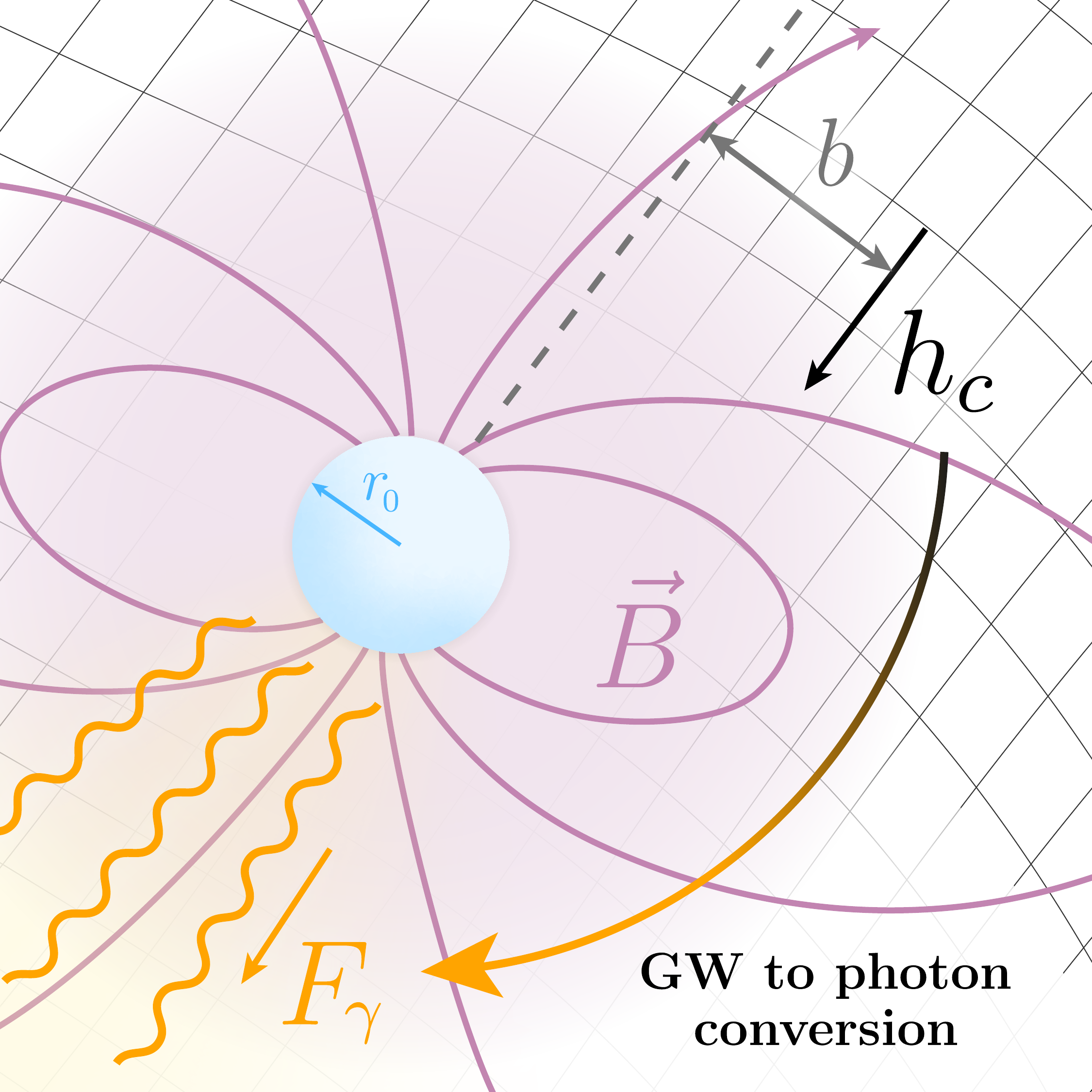}
    \hspace{0.05\textwidth}
    \includegraphics[width = 0.45\textwidth]{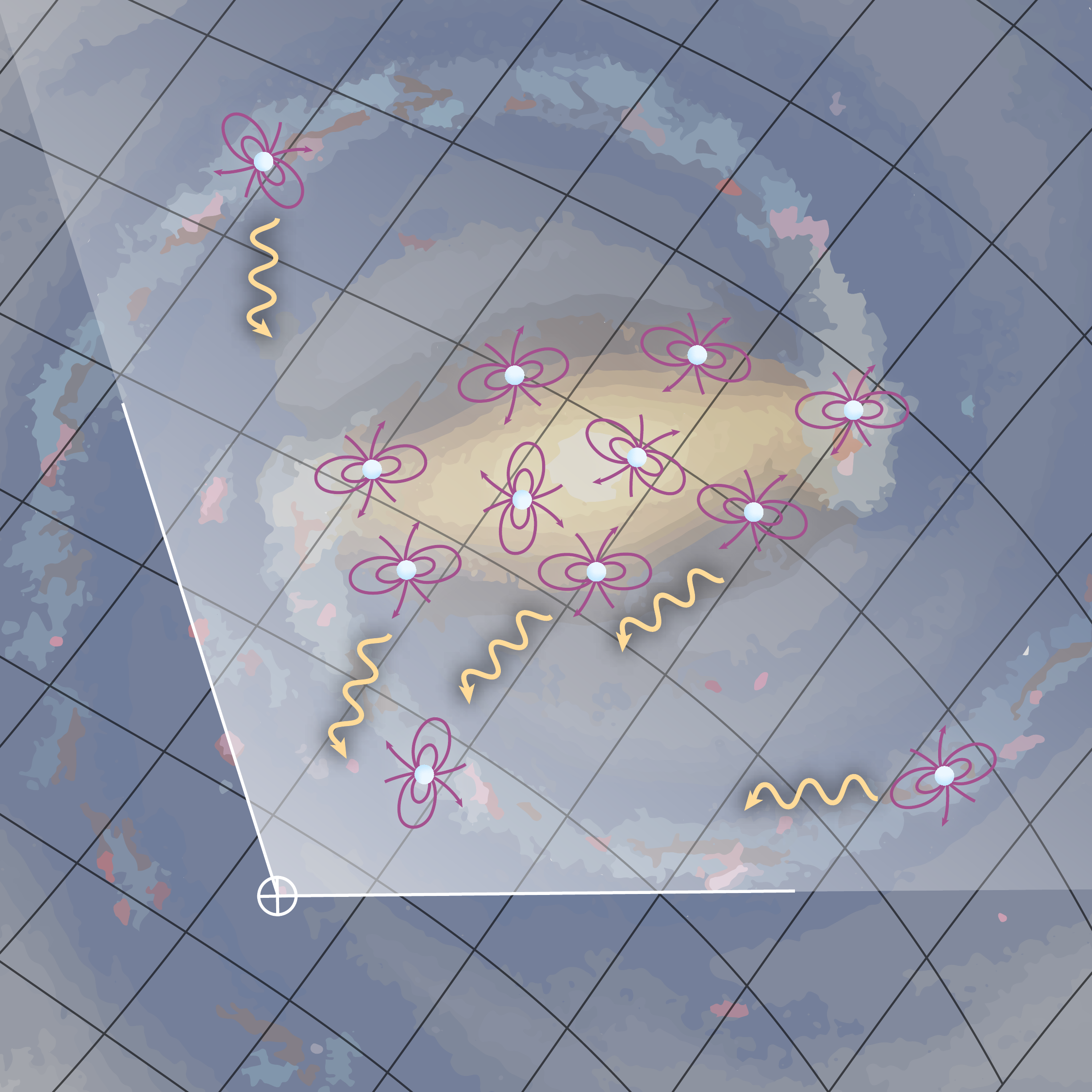}
    \caption{GWs with a characteristic strain $h_c$ can be converted into a flux of photons $F_\gamma$ inside a neutron star magnetic field $\Vec{B}$, through the so-called inverse Gertsenshtein effect. Here, $b$ is the impact factor of the graviton trajectory. The whole population of neutron stars inside the galaxy produces a photon flux on top of the known galactic emission from the graviton-photon conversion of a SWGB.}
    \label{fig: Cartoon NS}
\end{figure}
One of the main reasons that makes astrophysical environments particularly suitable for observing the Gertsenshtein effect lies in the natural presence of strong magnetic fields (which enhance the graviton-photon conversion) impossible to produce in the laboratory. Among the most extreme examples, neutron stars (NSs) are known to possess a surface magnetic field around $B \approx 10^{13}\, \rm Gauss$ \cite{Faucher-Giguere:2005dxp, Popov:2009jn} and, therefore, appear to be one of the most efficient places to convert gravitons into photons. Furthermore, NS magnetospheres are relatively well understood and allow for reliable predictions. In the presence of a SWGB the ensemble of NSs present in the Milky Way—estimated to be around $10^9$ NSs \cite{Ofek:2009wt, Sartore:2009wn}—would then induce a photon flux at the same frequency as the SGWB, as illustrated in Fig.~\ref{fig: Cartoon NS}. The galactic NS population therefore behaves as a powerful tool to probe the presence of a high frequency SGWB.\\
Motivated by the last statement, we calculate in this work the radiated photon flux from the conversion of a high frequency SGWB in the magnetosphere of the ensemble of NSs within the Milky Way (see Ref.~\cite{Safdi:2018oeu} for a similar study in the axion case). Using recent models for the NS magnetic field, spin period, and spatial distribution~\cite{Faucher-Giguere:2005dxp,Popov:2009jn}, we 
compare this 
photon flux to the observations conducted by several telescopes (see Ref.~\cite{Hill:2018trh} for a summary). Finally we extract competitive constraints on the amount of GWs in a large frequency range, from $\sim 10^8\, \rm Hz$ to $\sim 10^{25}\, \rm Hz$.

In the following, we first start in Sec.~\ref{Sec: Conversion Probability} by recalling the inverse Gertsenshtein effect in the specific context of NS magnetospheres. We describe in Sec.~\ref{Sec: Flux vs observation} the expected induced photon flux from the galactic NSs and detail the method used to extract constraints on the GW abundance. Finally, we show our results in Sec.~\ref{SubSec: upper limits} and conclude in Sec.~\ref{Sec: Conclusion}. The specific models for the NS distributions are discussed in App.~\ref{App: Neutron Star Distribution}.

\section{Graviton-Photon Conversion in Neutron Stars}\label{Sec: Conversion Probability}

The GW-to-photons conversion is described by the inverse Gertsenshtein effect which is the result of the coupling between gravity and electromagnetism via the Lagrangian term $\mathcal{L} \supset F_{\mu \nu} F^{\mu \nu} = g^{\rho \mu} g^{\sigma \nu}  F_{\rho \sigma} F_{\mu \nu}$. Note that we are working in natural units and using the $\eta_{\mu \nu} =$ diag$(1,-1,-1,-1)$ metric signature. Considering the GW (and then the photons) propagating in a direction perpendicular to the magnetic field, one can derive the leading order equation of motion for the gravitons and photons polarizations. Assuming a homogeneous magnetic field, that reads~\cite{Raffelt:1987im, Liu:2023mll}
\begin{align}
\left(\omega-i \partial_n+\left(\begin{array}{cccc}
\Delta_{\perp} & \Delta_M & \Delta_R & 0 \\
\Delta_M & 0 & 0 & 0 \\
\Delta_R & 0 & \Delta_{\|} & \Delta_M \\
0 & 0 & \Delta_M & 0
\end{array}\right)\right)\left(\begin{array}{c}
A_{\perp} \\
h_{+} \\
A_{\parallel} \\
h_{\times}
\end{array}\right)=0 .
\label{eq:equation-of-motion}
\end{align}
The GW polarizations are $h_\times$ and $h_+$, while $A_{\parallel}$ and $A_{\perp}$ are respectively the parallel and perpendicular photon polarizations to the magnetic field. 
 % The term $\Delta_R$ is proportional to $\Vec{B}-B_t \Vec{e}_t$ and is negligible compared to the other contributions since we are approximating the GW to be transverse to $\Vec{B}$ \TB{Why?}. We defined $\Vec{e}_t$ and $B_t$ as the unit vector in the transverse direction and the magnitude of the transverse component of the magnetic field. Meaning the component transverse to the direction of motion of the incoming GWs. 
As we will discuss in the following, we approximate the GW to be transverse to the magnetic field $\Vec{B}$.
Defining $B_t$ as the component transverse to the direction of motion of the incoming GW, and $\vec{e}_t$ as the unit vector in this transverse direction, then this means that the longitudinal component $\Vec{B}-B_t \Vec{e}_t$ is negligible. Since $\Delta_R$ is proportional to this longitudinal component, then we can assume $\Delta_R \approx 0$ and treat Eq.~\eqref{eq:equation-of-motion} as two independent equations of motion \cite{Liu:2023mll,Ito:2023nkq} that read 
\begin{align}
\left(\omega-i \partial_n+\left(\begin{array}{cc}
\Delta_{\|(\perp)} & \Delta_{\mathrm{M}} \\
\Delta_{\mathrm{M}} & 0
\end{array}\right)\right)\left(\begin{array}{c}
A_{\|(\perp)} \\
h_{\times(+)}
\end{array}\right)=0\, .
\end{align}
Notice that $\times(+)$ GW polarizations will be converted into $\|(\perp)$ photon polarizations.
The graviton-photon mixing is $\Delta_{\mathrm{M}}=\frac{1}{2} \kappa B_t$ with $\kappa=(16 \pi G)^{1/2}$. In the presence of the external magnetic field $B_t$, an effective mass $\Delta_{\|(\perp)}$ is induced for the photons, 
\begin{align}
\begin{aligned}
\Delta_{\| (\perp)} & =\Delta_{\mathrm{pla},\parallel(\perp)}+\Delta_{\mathrm{vac}, \|(\perp)}
\end{aligned}
\end{align}
Each term describes a different physical effect: $\Delta_{\mathrm{pla},\parallel}= -\frac{m_{\mathrm{pla}}^2}{2 \omega}$ and $\Delta_{\mathrm{pla},\perp}= -\frac{m_{\rm{pla}}^2 \omega^2}{2\omega (\omega^2 - \omega_c^2)}$ are the plasma corrections to the photon mass, with $\omega_c\approx 10^{19}\left(\frac{B}{10^{12}{\rm G} }\right) \,\mathrm{Hz}$ \cite{Raffelt:1987im}. The difference between the two polarizations comes from the Cotton-Mouton effect \cite{Ejlli:2018hke,Ejlli:2018ucq,Ito:2023fcr}. $\Delta_{\mathrm{vac}, \|(\perp)}=7(4) \frac{\alpha \omega}{90 \pi}\left(\frac{B_t}{B_c}\right)^2$ is the QED vacuum correction. We defined $m_{\mathrm{pla}}^2=\frac{4 \pi \alpha  n_c}{m_c}$, $B_c=\frac{m_e^2}{e}$ and $\omega = 2 \pi f$, with $f$ the graviton/photon frequency. Moreover, $\alpha$ and $e$ are the QED fine structure constant and the electron charge. Finally, $n_c$ and $m_c$ are the number density and invariant mass of the particles in the plasma. In NSs we have $n_c \approx n_e$ and $m_c \approx m_e$, the electron number density and the electron mass, respectively.\\
In NSs, the magnetic field is not homogeneous. Still, the previous system of equation can be solved perturbatively in the case of an inhomogeneous magnetic field (see Ref.~\cite{Raffelt:1987im}). With this, the total GW-photon conversion probability can be extracted for the $\|$ ($\perp$) polarization and is given by~\cite{Raffelt:1987im,Chen:1994ch}
%
%\VD{. $\Delta_{\mathrm{CM}} =\frac{m_{\rm{pla}}^2 \omega_c^2}{2\omega (\omega^2 - \omega_c)}$ describes the Cotton-Mouton birefringence effect, with $\omega_c\approx 10^{19}\left(\frac{B}{10^{12}{\rm G} }\right)$ \cite{Raffelt:1987im}.}
%
%The effective photon mass in the plasma is given by $\Delta_\gamma \approx \Delta_{\mathrm{vac}}+\Delta_{\mathrm{pla}}$. \FC{\st{The difference between the two GW polarisation modes is negligible } \cite{Raffelt:1987im} [check reference] 2305.13984 says the opposite}. $\Delta_{\mathrm{vac}}$ is the QED correction to the mass and $\Delta_{\mathrm{pla}}$ is the plasma correction, they are given respectively by
% \begin{align}
% \Delta_{\mathrm{vac}}&=\frac{14 \pi \alpha f}{90 \pi}\left(\frac{B_t}{B_c
% }\right)^2,\\
% \Delta_{\mathrm{pla}}&=-\frac{m_{\mathrm{pla}}^2}{4 \pi f
% },
% \end{align}
%
\begin{align}
P_{\|(\perp)}(f)=\left|\int_{\ell_0}^{\ell_1} \mathrm{d} \ell\, \Delta_{\mathrm{M}}(\ell) \exp \left\{-i \int_{\ell_0}^{\ell} \mathrm{d} \ell^{\prime}\, \Delta_{\|(\perp)}\left(\ell^{\prime}\right)\right\}\right|^2 \, , 
\label{eq.proabibility}
\end{align}
considering a generic GW path in the magnetic field between $\ell_0$ and $\ell_1$\footnote{In what follows, we assume that the typical wavelength of the GW is much smaller than the NS magnetosphere. This, in addition to the large occupation number, allows us to describe GWs as made of individual gravitons with defined trajectories. A more precise study of the problem using a wave description of the system is left for future works.}.\\ 
In the specific case of NSs, the magnetic field extends from its surface $r_0$ to the end of the magnetosphere $R\approx T/(2\pi)$ \cite{Ito:2023fcr}, with $T$ the rotation period. Assuming equipartition and averaging over the directions of the magnetic field we parametrize the component of the magnetic field perpendicular to the GW trajectory in the NS magnetosphere as~\cite{Ito:2023fcr}
\begin{align}
B_t=\sqrt{\frac{2}{3}} B_0\left(\frac{r}{r_0}\right)^{-3},
\end{align}
where $B_0$ is the magnetic field at the NS surface, $r$ the distance to the NS center and $r_0$ the radius of the NS. We consider the magnetic field as a dipole aligned with the rotation axis. This leads to a reliable upper bound on the GW strain, since a dipole provides the minimum amount of GW-photon conversion.\\
Finally, the electron number density $n_e$ in the magnetosphere can be described in a Goldreich-Julian model \cite{Goldreich:1969sb,Ito:2023fcr}
%Finally, the electron number density considered is ~\cite{Ito:2023fcr}
\begin{align}
n_e (r)=7 \times 10^{-2}\left(\frac{1\, \mathrm{s}}{T}\right)\left(\frac{B_t(r)}{1\, \mathrm{G}}\right) \mathrm{cm}^{-3} \, .
\end{align}
In what follows we first derive an analytic expression for the conversion probability of gravitons with radial trajectories inside the magnetosphere, and then we compute it for arbitrary trajectories.
%\TB{\st{Note that} Eq.\eqref{eq:equation-of-motion} \st{implies a different conversion probability depending on the graviton trajectory inside the magnetosphere.}}
%In what follows we first derive an analytic expression for the conversion probability of gravitons with radial trajectories. \TB{\st{Which can be converted along the path from the NS surface to the end of the magnetosphere.}} Secondly, we derive the conversion probability for gravitons with arbitrary trajectories inside the magnetosphere.

\subsection{Radial Trajectory}
For radial trajectories throughout the magnetosphere, the gravitons would propagate from $r_0$ to $R$. Since all relevant quantities are spherically symmetric, Eq.~\eqref{eq:equation-of-motion} simply becomes
\begin{align}
\begin{split}
P_{0,\|(\perp)}(f)&=2\left|\int_{r_0}^{R} \mathrm{d}r\, \Delta_{\mathrm{M}}(r) \exp \left\{-i \int_{r_0}^{r} \mathrm{d} r^{\prime}\, \Delta_{\|(\perp)}\left(r^{\prime}\right)\right\}\right|^2 \, ,
\end{split}
\label{eq.proabibility}
\end{align}
where the factor $2$ arises since the graviton is actually traveling twice along this path (once when entering the magnetosphere and once when exiting it).\\
While this integral will be solved numerically in the next sections, we derive here analytical expressions in the limit of high and low frequencies. 
%\st{In Fig.2 we present the result of the previous integral for a benchmark NS with $B_0 = 10^8$ Gauss, $T_{\mathrm{NS}} = 0.1$ s and $r_0=10$ km}}. 
To this point, we choose as the upper limit of the integral $R \rightarrow \infty$. In fact, the magnetic field decays cubically 
%\TB{\st{quadratically} with $1/r^{3}$} 
with the distance and the unphysical contribution outside the magnetosphere is negligible. 

Let us start with the parallel polarization. Since $\Delta_{\mathrm{pla},\parallel}$ and $\Delta_{\mathrm{vac},\parallel}$ scale as $1/f$ and $f$ respectively, we have that at high frequencies we can approximate $\Delta_{\|} \approx \Delta_{\mathrm{vac},\|}$ and low frequencies $\Delta_{\|} \approx \Delta_{\mathrm{pla},\parallel}$. Thus, at high frequencies 
\begin{align}\label{eq:high-f-prob}
      P_{0,\|}^{\mathrm{high-f}} \approx  \frac{2 (\Delta_{\rm M}(r_0) r_0)^2}{5^{6/5} (\Delta_{\rm{vac,\parallel}}(r_0)r_0)^{4/5}}
      \left| \Gamma\left( \frac{2}{5} \right)  - \Gamma \left(\frac{2}{5}, -\frac{ i \Delta_{\rm{vac,\parallel}}(r_0) r_0}{5} \right)\right|^2,\nonumber
\end{align} 
% \begin{align}\label{Eq: Large freq}
%       P_{0,\|}^{\mathrm{high-f}}& \approx  0.3  \kappa^2 \left( \frac{m_e^4}{e^2 \alpha f  }  \right)^{4/5} B_0^{2/5} r_0^{6/5} \\&\times \left( \;\mathrm{Im}\left[\Gamma \left(\frac{2}{5}, -i x \right)\right]^2 + \left( \;  \Gamma\left( \frac{2}{5} \right)      - \; \mathrm{Re}\left[ 
%  \Gamma \left( \frac{2}{5}, - i x \right)\right] \right)^2  \right),\nonumber
% \end{align} 
% with 
% \begin{align}
%     x = \frac{14  \alpha e^2 B_0^2 r_0 f}{675 m_e^4},
% \end{align}
% $$x=1/5*\Delta_{\rm vac.\|
% }(r_0) FRANCESCO CHANGE IT!$$
where $\Gamma(x,y)$ is the incomplete Gamma function. Similar expression has been derived for instance in Ref.\cite{Liu:2023mll}.\\
At low frequencies,
% \begin{align}
%    P_{0,\|}^{\mathrm{low-f}}&  \approx 5.7 \times 10^{-4}\,  \frac{  \kappa^2 m_e^2 T^2}{ \alpha^2} f^2  \times \left( 1- \cos\left[ \frac{17 \alpha B_0 r_0}{m_e f T}\right] \right).
% \end{align}
\begin{align}
   P_{0,\|}^{\mathrm{low-f}}&  \approx  \frac{2\Delta_{\rm M}^2(r_0)}{\Delta^2_{\rm{pla},\|}(r_0)} \left| \exp{ \left(\frac{i  \Delta_{\rm{pla},\|}(r_0) r_0}{2}\right)} -1   \right|^2.
\end{align}
The high (low) frequency regime is defined for frequencies in general much higher (lower) than the transition frequency. The transition frequency can be found by $|\Delta_{\mathrm{pla},\parallel}|\approx|\Delta_{\mathrm{vac},\parallel}|$, 
\begin{equation}
    f_{\mathrm{tra}} \approx 6.5 \times \frac{m_e^{3/2}}{e}\sqrt{\frac{2 \pi}{B_0 T}} \, .
\end{equation}
For frequencies around $f_{\mathrm{tra}}$, there is a region where $\Delta_{\mathrm{pla},\parallel}$ and $\Delta_{\mathrm{vac},\parallel}$ compensate, and thus the photon effective mass vanishes. This region is close to the NS surface.
When this mass is small enough, the graviton-photon conversion satisfies both momentum and energy conservation and is enhanced. The maximum of the conversion probability would therefore be around $f_{\mathrm{tra}}$.
A good approximation of the maximum conversion probability is given by 
\begin{equation}
    P^{\rm max}_{0,\|} \approx P_{0,\|}^{\mathrm{high-f}}(f_{\mathrm{tra}}).
\end{equation}
%\begin{equation}
%    P_0^{\mathrm{plateau}} \approx \frac{\kappa^2 %B_0^2r_0^2}{12}\, .
%\end{equation}
%At $f_\mathrm{tra}$ the plasma and vacuum contributions cancel and the effective photon mass is minimal.
%Around $f_{\mathrm{tra}}$ the conversion probability is the highest. \FC{\st{Indeed, since the plasma and vacuum contributions cancel each other, the effective photon mass does the same}In fact the effective photon mass here is minimal because of a cancellation between the plasma and vacuum contributions}. 
%When the effective mass is small enough, the graviton-photon conversion satisfies both momentum and energy conservation and is therefore enhanced.
In contrast, for much larger (smaller) frequencies, the effective photon mass provided by $\Delta_{\rm vac,\parallel}$ ($\Delta_{\rm pla,\parallel}$) decreases significantly the conversion probability \footnote{In practice, for much larger frequencies than $f_{\mathrm{tra}}$ there is still a region where $\Delta_{\mathrm{pla},\parallel}$ and $\Delta_{\mathrm{vac},\parallel}$ compensate. However, as the frequency increases, this one is pushed to larger radius and is suppressed by the decaying magnetic field.}\footnote{This is actually a general statement: the larger the effective mass, the lower the conversion probability.}.

\begin{figure}[h]
	\centering
	\begin{subfigure}{.5\textwidth}
		\centering
		\includegraphics[scale=0.52]{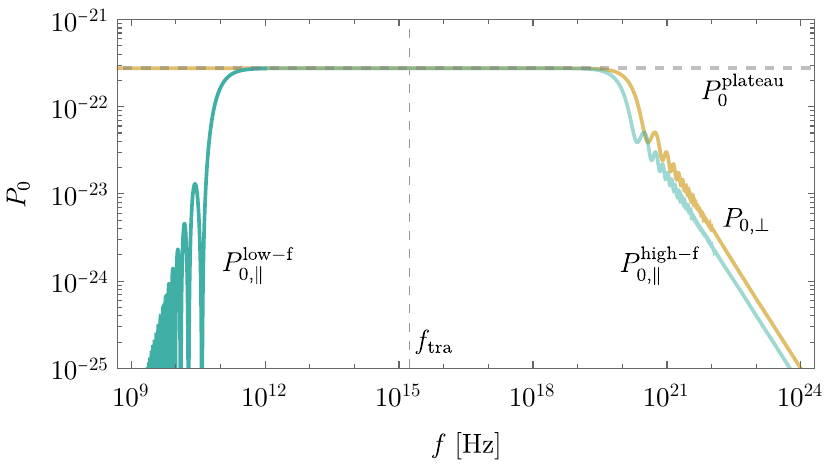}
		%	\caption{A subfigure}
		\label{fig:sub1}
	\end{subfigure}%
	\begin{subfigure}{.5\textwidth}
		\centering
		\includegraphics[scale=0.52]{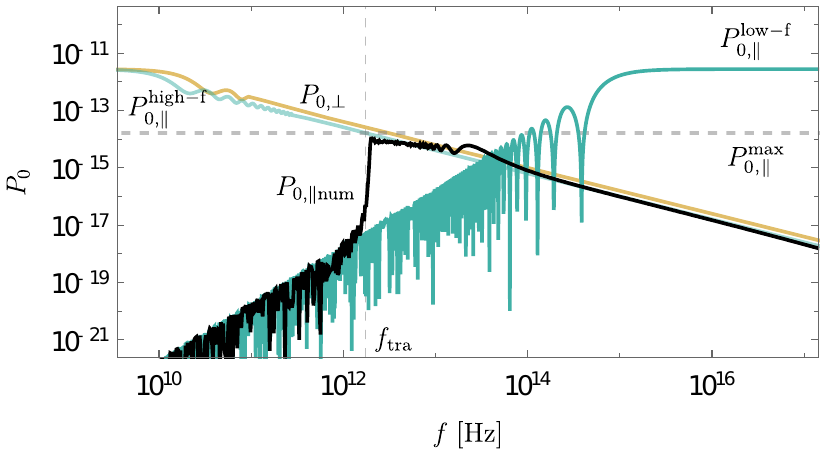}
		%	\caption{A subfigure}
		%	
	\end{subfigure}
    \caption{Conversion probability as a function of the frequency for an NS with $B_0 = 10^8$ Gauss, $T = 0.1$ s, and $r_0=10$ km (left) and for $B_0 = 10^{13}$ Gauss, $T = 1$ s, and $r_0=10$ km (right). The curves respectively in dark and light green are the analytical approximation at low frequencies and high frequencies for the parallel polarization. The yellow curve is the analytical approximation for the perpendicular polarization which matches the numerical result. The dashed gray line is the maximum conversion probability approximation for the parallel polarization and the black dashed line is $f_{tra}$. Finally, the black curve in the right panel is the numerical result for the parallel polarization.}
    \label{fig:probability}
\end{figure}

For the perpendicular polarization, we notice that in the typical NS magnetic fields $\Delta_{\rm{pla},\perp}\approx 0$ due to a $\omega_c\approx 10^{19}\left(\frac{B}{10^{12}{\rm G} }\right) \,\mathrm{Hz}$ suppression. This means that $\Delta_\perp \approx \Delta_{\mathrm{vac},\perp}$ and we have a continuation of the high-frequency behavior as described in Eq.~\eqref{eq:high-f-prob} even at low frequencies. The perpendicular polarization conversion probability is simply given by Eq. \eqref{eq:high-f-prob} with $\Delta_{\rm{vac,\parallel}}(r_0) \rightarrow \Delta_{\rm{vac,\perp}}(r_0)$.
% \begin{align}\label{Eq: Large freq,parallel}
%       P_{0,\perp}& \approx  0.46 \kappa^2 \left( \frac{m_e^4}{e^2 \alpha f  }  \right)^{4/5} B_0^{2/5} r_0^{6/5} \\&\times \left( \;\mathrm{Im}\left[\Gamma \left(\frac{2}{5}, -i y \right)\right]^2 + \left( \;  \Gamma\left( \frac{2}{5} \right)      - \; \mathrm{Re}\left[ 
%  \Gamma \left( \frac{2}{5}, - i y \right)\right] \right)^2  \right),\nonumber
% \end{align} 
% \begin{align}\label{Eq: Large freq,parallel}
%       P_{0,\perp}& \approx  \frac{2 \Delta_M(r_0)}{5^{6/5} \Delta_{\rm{vac,\perp}}(r_0)^{4/5}}
%       \left( \;\mathrm{Im}\left[\Gamma \left(\frac{2}{5}, -i \frac{\Delta_{\rm{vac,\perp}}(r_0)}{5 r_0^5} \right)\right]^2 + \left( \;  \Gamma\left( \frac{2}{5} \right)      - \; \mathrm{Re}\left[ 
%  \Gamma \left( \frac{2}{5}, - i \frac{\Delta_{\rm{vac,\perp}}(r_0)}{5 r_0^5} \right)\right] \right)^2  \right),\nonumber
% \end{align} 
% \begin{align}\label{Eq: Large freq,parallel}
%       P_{0,\perp}& \approx  \frac{2 (\Delta_M(r_0) r_0)^2}{5^{6/5} (\Delta_{\rm{vac,\perp}}(r_0)r_0)^{4/5}}
%       \left| \Gamma\left( \frac{2}{5} \right)  - \Gamma \left(\frac{2}{5}, -\frac{ i \Delta_{\rm{vac,\perp}}(r_0) r_0}{5} \right)\right|^2,\nonumber
% \end{align} 
% with 
% \begin{align}
%     y = \frac{8  \alpha e^2 B_0^2 r_0 f}{675 m_e^4},
% \end{align}
%
As we will see, this will induce that at low frequencies gravitons with $+$ polarization have a much higher chance of being converted into $\|$ photons, as opposed to $\times$ gravitons.
%Therefore at low frequencies, the perpendicular contribution strongly dominates the probability.

In Fig.~\ref{fig:probability} we present the analytical results in dark green (high-frequency) and light green (low-frequency) for the parallel polarizations and in yellow for the perpendicular one. The left(right) panel corresponds to a surface magnetic field of $10^{8}$ ($10^{13}$) Gauss.\\
\noindent For the $\parallel$ polarization, we find good agreements with the numerical solution of Eq.~\eqref{eq.proabibility} (black curve in the right panel) for frequencies much larger and much smaller than $f_{\rm tra}$. For frequencies around $f_{\rm tra}$, the conversion probability is enhanced due to a region on the path of the graviton where $\Delta_{\mathrm{pla},\parallel}$ and $\Delta_{\mathrm{vac},\parallel}$ compensate. We denote by $P^{\rm max}_{0,\|}$ the maximum conversion probability for the $\parallel$ polarization. Since the plasma(vaccuum) effect scales as $\sim 1/f$($\sim f$), at lower(larger) frequencies the effective mass increases significantly. From the same arguments, it explains why the conversion probability drops quickly in these regions.\\
\noindent For the $\perp$ polarization, the conversion probability is simply given by Eq.~\eqref{eq:high-f-prob} (replacing $\Delta_{\mathrm{vac},\parallel}$ by $\Delta_{\mathrm{vac},\perp}$). Since $\Delta_{\mathrm{vac},\perp} \sim f $, the effective photon becomes smaller as we lower the frequency. Therefore, the conversion probability increases as the frequency gets lower, until it reaches a maximum when the effective mass becomes negligible. The numerical results match the analytical approximation.
 
%%%   SOVLED I THINK
%[define this better, maybe using r0?, \TB{We can define it in the limit where the plateau shrinks the most. Or get it analytically from the maximum of the function, which coincides with the plateau. Or equate $\Delta_{pla} = \Delta_{vac}$ and take it from there}]
%%%
%

\subsection{Generic Trajectory}

Most gravitons crossing the magnetosphere are not radial, but have a linear trajectory with a particular impact parameter $b$, as shown in Fig.~\ref{fig: Cartoon NS}.
Such path can be parametrized if we align the $y$ axis to the graviton trajectory, and thus Eq.~\eqref{eq:equation-of-motion} becomes 
\begin{align}
\begin{split}
P_{\|(\perp)}(f,b)&=2 \left|\int_{y_{\rm min}}^{y_{\rm max}} \mathrm{d}y \, \Delta_{\mathrm{M}}(y) \exp \left\{-i \int_{y_{\rm min}}^{y} \mathrm{d} y^{\prime}\, \Delta_{\|(\perp)}\left(y^{\prime}\right)\right\}\right|^2 \,.
\end{split}
\label{eq.proabibility_arbitrary}
\end{align}
Here $y_{\rm max} = \sqrt{R^2-b^2}$ and $y_{\rm min} = 0 $ if $b>r_0$ (the graviton is not crossing the NS) and $y_{\rm min} = \sqrt{r_0^2-b^2}$  if $b<r_0$ (the graviton is crossing the NS). Similarly to the radial trajectory, the factor 2 arises from symmetry.

We can  gain some intuition on what this probability should look like. We have seen in the previous section that the conversion is efficient when the effective photon mass is small (i.e. $\Delta_{\|(\perp)}\approx 0$). Since $\vec{B}$ decays as $1/r^3$, $\Delta_{\|(\perp)}$ also decreases as the radius increases. Thus, for each frequency, there exists a radius further from which gravitons can convert efficiently. However, this ``sweet'' conversion spot does not extend much further because $\Delta_{\rm M}$ drops quickly too. Hence due to the necessity of having both a large magnetic field and a small photon mass, the efficient conversion region is limited to a thin shell of critical radius $b_c(f)$ right after the effective photon mass becomes negligible. \\ % \TB{We will call $b_c$ the radius of this shell.} \\
As long as the graviton crosses this shell (i.e. $b<b_c(f)$), the probability of conversion remains constant, since this contribution strongly dominates the integral \eqref{eq.proabibility_arbitrary}. In particular, it matches the probability calculated in the previous section: $P_{0,\|(\perp)}(f)$.
If the impact parameter $b$ becomes large enough so that the graviton does not cross  the shell (i.e. $b>b_c(f)$), then we should expect for the conversion probability to be suppressed due to the decaying magnetic field. In particular, an analytical computation shows a $1/b^4$ suppression on the probability. 
\begin{figure}[t!]
    \centering
    \includegraphics[width=1\textwidth]{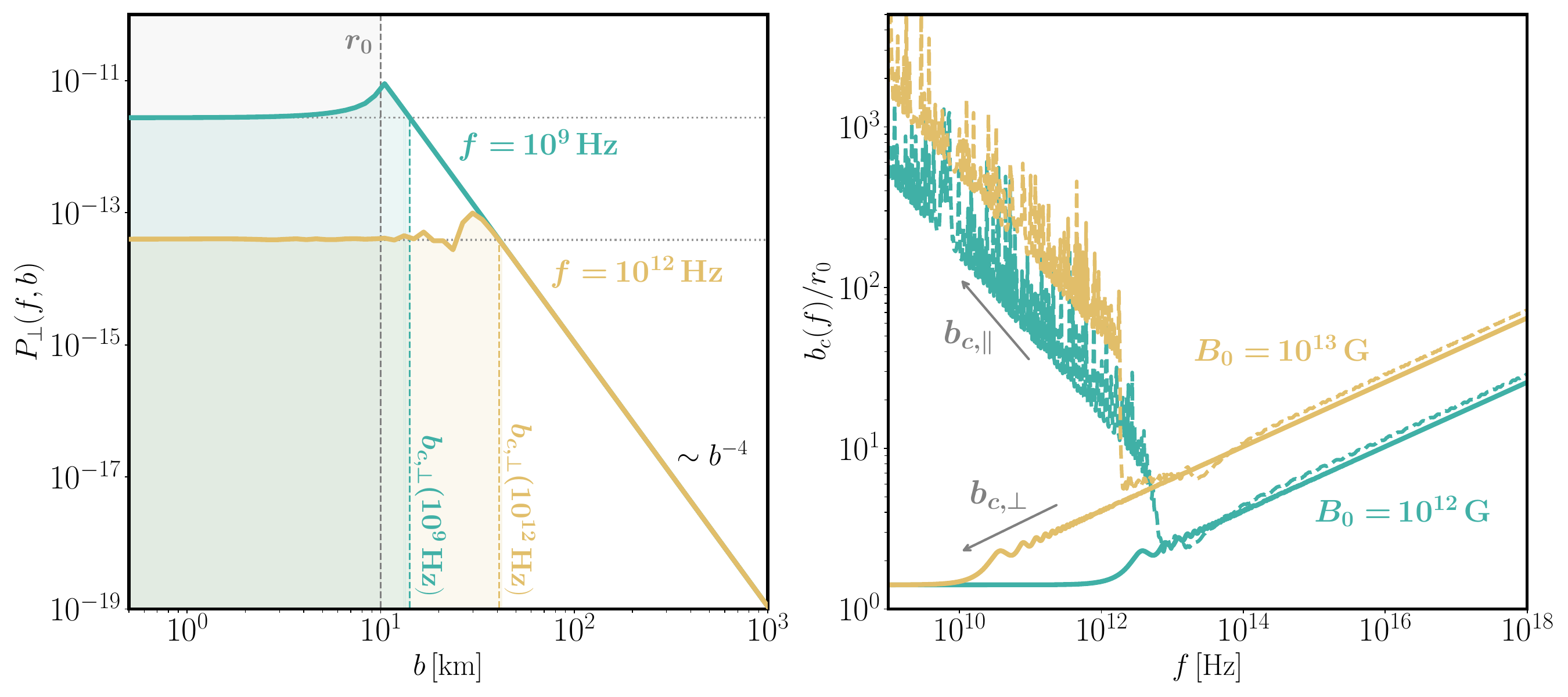}
    \caption{\textit{Left}: Conversion probability for the perpendicular polarization as a function of the impact parameter. Green and yellow lines show the probability for $f=10^{9} \, \rm Hz$ and  $f=10^{12} \, \rm Hz$, respectively. Vertical and horizontal dashed lines show the value of $b_c(f)$ and $P_0^{\rm plateau}$ for each frequency, respectively. The magnetic field is set to $B_0=10^{13}\, \rm G$. \textit{Right}: Critical impact parameter as a function of the frequency. Green and yellow lines show $B_0=10^{12}\, \rm G$ and $B_0=10^{13}\, \rm G$, respectively. Solid and dashed lines show the perpendicular and parallel polarizations, respectively. In both plots, the NS spin period and radius has been taken to be $T = 1$ s, and $r_0=10$ km.}
    \label{fig: arbitrary trajectory}
\end{figure}

A numerical result for the probability \eqref{eq.proabibility_arbitrary} as a function of $b$ is shown in the left panel of Fig.~\ref{fig: arbitrary trajectory}.
In the right panel, we show the critical impact parameter $b_c(f)$. Since the photon effective mass depends on both the magnetic field and the polarization, so does $b_c(f)$. We make this explicit by writing $b_{c, \parallel}(f,B_0)$ and $b_{c, \perp}(f,B_0)$. They differ as follows:
\begin{itemize}
    \item For the $\parallel$ polarization, the effective photon mass is given by $\Delta_{\rm vac, \parallel}+\Delta_{\rm pla,\parallel}$. Since $\Delta_{\rm pla, \parallel}\propto 1/f$, the $\parallel$ polarization has a larger effective photon mass at small frequencies. Therefore, the critical impact paramater $b_{c,\parallel}(f,B_0)$ increases at low $f$. 
    \item For the $\perp$ polarization, $\Delta_{\rm vac, \perp}$ is suppressed at small frequencies, and thus $b_{c,\perp}(f)$ decreases with $f$. 
\end{itemize}
%In particular, the effective photon mass is given by $\Delta_{\rm vac, \parallel}+\Delta_{\rm pla,\parallel}$ and $\Delta_{\rm vac, \perp}$ for the $\parallel$ and $\perp$ polarization respectively. 
%On the one hand, $\Delta_{\rm vac, \perp}$ is suppressed at small frequencies, and thus $b_{c,\perp}(f)$ decreases with $f$. 
%the $\perp$ polarization has a smaller critical impact parameter as we decrease the frequency. 
%On the other hand, $\Delta_{\rm pla}$ grows with $1/f$,
%as the frequency decreases, 
%so that the $\parallel$ polarization has a larger effective photon mass at small frequencies. For this reason, the critical impact paramater $b_{c,\parallel}(f,B_0)$ increases at low $f$. 
%\TB{Maybe needs rephrasing...}
Finally, the effective photon mass grows with the surface magnetic field $B_0$, so $b_c(f)$ increases with it too. %  so that we find that increasing this one leads to a larger critical impact parameter.\\

In what will follow, we consider that, for a given frequency, only gravitons  with $b<b_{c,\parallel (\perp)}(f,B_0)$ are converted. For those gravitons the conversion probability is the one calculated under the radial approximation. Explicitly,
\begin{equation}
    P_{\|(\perp)}(f,b,B_0) \approx \left\{
    \begin{array}{ll}
        P_{0,\|(\perp)}(f,B_0)\ \ & \mbox{if}\ \  b<b_{c,\parallel (\perp)}(f,B_0)\\
        0 & \mbox{if}\ \ b>b_{c,\parallel (\perp)}(f,B_0)
    \end{array}
\right.\, .
\end{equation}

%\begin{itemize}
    %\item Take the radial probability of conversion $P_{r_0\to R}$
    %\item Assume that this probability holds for all trajectories coming from inside the neutron star (not magnetosphere). This will put a conservative bound.
    %\item Assume an isotropic diffuse GW background flux and compute the flux from one neutron star.
    %\item Integrate for the galactic distribution of neutron stars, both in position and in NS properties (magnetic field and period).
    %\item Optional: In this manner we can obtain a directional flux from the galaxy. (then what do we do with this? if we do not do anything, then it should go to the appendix). We keep one of the figures, as an illustration of the angular distribution of the flux.
    %\item In this manner we can obtain a frequency spectrum for the total photon flux. Add plot.
    %\item We can then compare this spectrum to the observed extragalactic spectrum. Add plot. This is an all-sky integrated flux, which could be improved by looking a certain region of the sky (e.g. GC)
%\end{itemize}

\section{Photon Flux from GW Conversion in Galactic Neutron Stars}\label{Sec: Flux vs observation}
As we have just seen, the magnetosphere of NSs is an interesting region for conversion of GWs to photons. 
Then, a diffuse and isotropic flux of HFGWs would be converted into a galactic photon flux, which would be strongly correlated with the galactic NS distribution. In this section we use the conversion probability just derived to compute the total photon flux from the conversion of a HFGW background on each of the NSs of the galaxy.

An isotropic stochastic GW background carries a flux of gravitons $F_{\rm gw}$ which can be written in terms of its spectral energy density $d\rho_{\rm gw}/df$ as~\cite{Ramazanov:2023nxz}
\begin{equation}
    \frac{\partial F_{\rm gw}}{\partial f \partial \Omega} = \frac{1}{4\pi} \frac{d \rho_{\rm gw}}{df}\, ,
\end{equation} where $\Omega$ is the solid angle. 
%, per solid angle and per frequency, given by ,
%
%\begin{equation}
%    \frac{\partial F_{\rm gw}}{\partial f \partial \Omega} = \frac{1}{4\pi} \frac{d \rho_{\rm gw}}{df},
%\end{equation}
%
%where $d\rho_{\rm gw}/df$ is the spectral energy density of the SGWB.\\
It follows that the outgoing flux of gravitons is
\begin{equation}
    \frac{\partial F_{\rm gw}}{\partial f} = \pi^2 M_{\rm pl}^2 f h_c^2,
\end{equation}
where $M_{\rm pl}$ is the reduced Planck mass, and we have integrated over half a sphere. We have introduced the characteristic strain $h_c$~\cite{Maggiore:2000gv},\\
\begin{equation}
    \frac{d\rho_{\rm gw}}{df} = 2\pi^2 h_c^2 M_{\rm pl}^2 f\, .
\end{equation}
Gravitons convert only if they cross the magnetosphere with an impact parameter $b<b_{\parallel (\perp)}(f,B_0)$. Then, the induced photon flux, $F_{\gamma}$, received on Earth is 
%$P_{0,\|(\perp)}(f)$
\begin{equation}\label{Eq: Single Pulsar Flux}
     \frac{\partial F_{\gamma}}{\partial f} = \pi^2 M_{\rm pl}^2 f h_c^2 \left[\left(\frac{b_{c,\|}(f,B_0)}{r}\right)^2 P_{0,\|}(f)+\left(\frac{b_{c,\perp}(f,B_0)}{r}\right)^2 P_{0,\perp}(f)\right],
\end{equation}
where $r$ is the distance of the NS to Earth.
% A factor $2$ arises from the fact that gravitons are actually crossing twice the magnetosphere. 
% I CHANGED A BIT THE PHRASING, I DON'T THINK WHAT ITO ET AL ARE DOING IS WRONG AND DON'T WANT TO MAKE THAT STATEMENT. I JUST BELIEVE THAT FORMULA CANNOT BE USED (AND THEY SAY THAT THEY DO NOT USE IT)
%This last expression is essentially the one obtained in Ref.~\cite{Ito:2023fcr} with the difference that we considered only the gravitons crossing the NS core leading to the factor $r_0$ at the numerator. \TB{Alternative: that we considered only the gravitons which go from }
%In the expression from~\cite{Ito:2023fcr}, all gravitons crossing the magnetosphere are considered to convert with the same conversion probability, leading to an overestimation of the induced photon flux on Earth.\\

NSs are distributed in the galaxy following a distribution $n_{\rm ns}(\boldsymbol{r})$. Since the conversion probability \eqref{eq.proabibility_arbitrary} depends on the NS magnetic field $B_0$ and rotation period $T$, we also need their distribution among galactic NSs. With respect to these, the galactic NS propulation is divided into two main subgroups: active and dead stars. While the former can emit a coherent radio pulsed emission and are detectable, the latter have already lost too much spin through such emission and are not detectable anymore. 
Thus, dead NSs have, in general, a longer period. Still, both have an active magnetosphere and must be taken into account (see App.~\ref{App: Neutron Star Distribution}). \\
From NS observations, different models for their evolution and properties have been developed. We choose two of them~\cite{Faucher-Giguere:2005dxp,Popov:2009jn}, which differ in the evolution of the magnetic field. While~\cite{Faucher-Giguere:2005dxp} has a constant magnetic field,~\cite{Popov:2009jn} accounts for its decay and predicts a weaker magnetosphere for dead NS. Although both models are consistent with data (see for instance Refs.~\cite{Manchester:2001fp,Edwards:2001de}), the former leads to stronger constraints and the latter to more conservative ones. We keep both in order to illustrate the sensitivity of our predictions to the models, which is the dominant uncertainty of our results. Then, each model predicts a different distribution for both variables. Independently of the model, the probability distribution for both variables inside of each subgroup is well represented by a log-normal distribution, and so we distinguish $P_{\rm act.}(B_0, T)$ for active NSs from $P_{\rm dead}(B_0, T)$ for dead NSs. It is estimated that 0.4\% of the galactic NS population are active, while 99.6\% are dead~\cite{Safdi:2018oeu}. Further details on the spatial and probability distributions can be found in App.~\ref{App: Neutron Star Distribution}.

All in all, the total induced photon flux from the galactic NSs, $F^{\rm galac.}_{\gamma}$, can be obtained by integrating over the whole galactic NS distributions, 
\begin{equation}\label{Eq: Galactic Flux}
\begin{split}
    \frac{\partial F^{\rm galac.}_{\gamma}}{\partial f}(f,h_c) = &\int \mathrm{d}^3\boldsymbol{r} \, n_{\rm ns}(\boldsymbol{r}) \int \mathrm{d}B_0\, \mathrm{d}T \, \Big( 0.004\times P_{\rm act.}(B_0,T)\\
    &+ 0.996 \times P_{\rm dead}(B_0,T) \Big) \frac{\partial F_{\gamma}}{\partial f}(B_0,T,\boldsymbol{r},f,h_c).
\end{split}
\end{equation}
\begin{figure}[t!]
    \centering
    \includegraphics[width=1\textwidth]{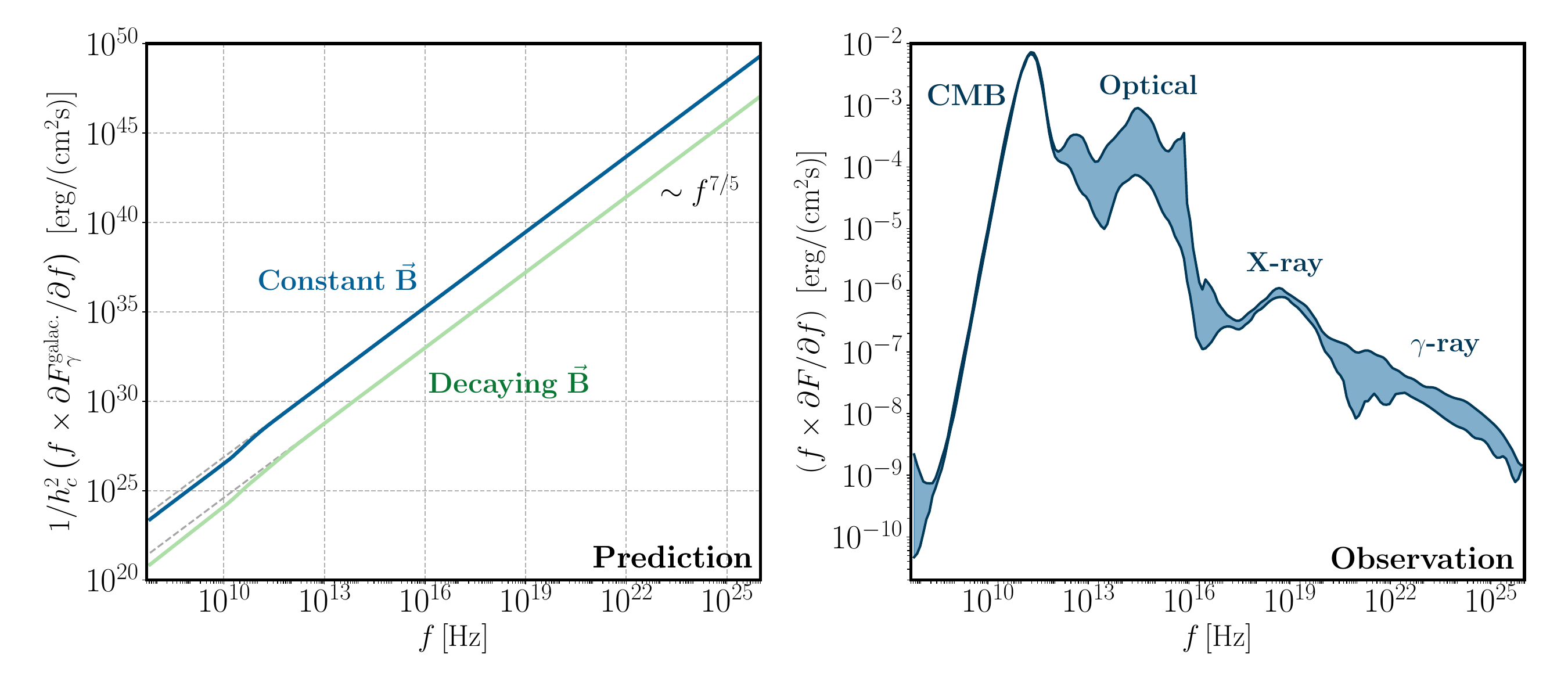}
    \caption{\textit{Left}: Predicted frequency spectrum of the total induced flux $\partial F^{\rm galac.}_{\gamma}/\partial f$, normalized by $1/h_c^2$. Blue and green lines show the results for a NS model with constant and decaying $\Vec{B}$, respectively. We set $r_0 = 10\, \mathrm{km}$. \textit{Right}: Frequency spectrum measured on Earth, adapted from Ref.~\cite{Hill:2018trh}. The thick band corresponds to the 95\% confidence interval. Data allows to put a bound on $h_c$ at each frequency.}
    \label{fig: Flux vs observation}
\end{figure}
%
% I COMMENT THIS BECAUSE IT MAKES THE PDF TOO HEAVY, I WILL UNCOMMENT IT LATER
\begin{figure}[t!]
    \centering
    \includegraphics[width=0.6\textwidth]{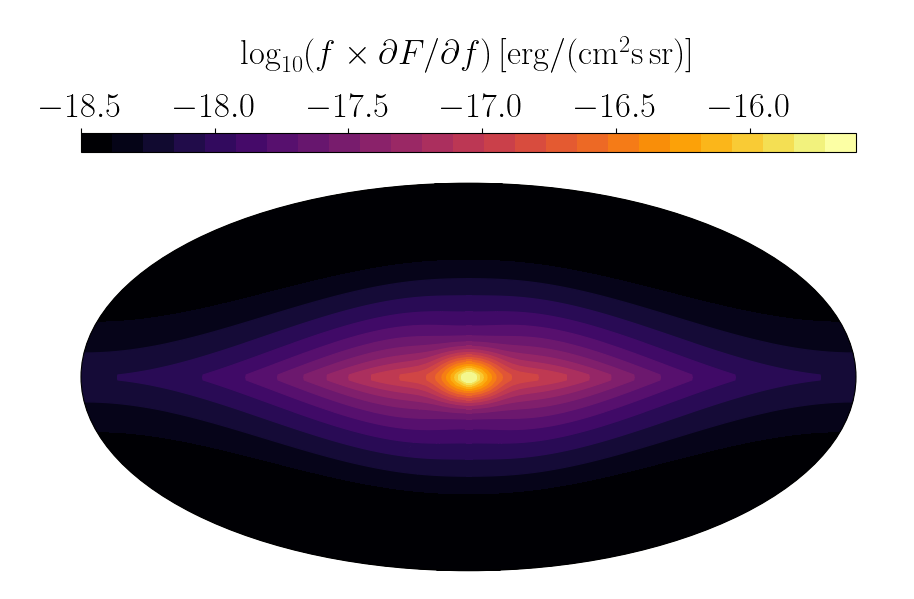}
    \caption{All sky map of the induced photon flux from the conversion of an isotropic SGWB with $h_c=10^{-25}$ and frequency $f = 10^{15} \rm Hz$. We considered a population of NSs with constant magnetic field. When considering the case of a decaying magnetic field, similar angular distribution is expected with a flux reduced by a factor $\mathcal{O}(10^{-3})$.}
    \label{fig: Angular Flux}
\end{figure}
In Fig.~\ref{fig: Flux vs observation} (left), we plot the frequency dependence of this flux, normalised as $h_c^{-2} f\, \partial F^{\rm galac.}_{\gamma}/\partial f$. Since $F^{\rm galac.}_{\gamma}(f,h_c)\propto h_c^{2}$, this is a quantity independent of the characteristic strain. 
At large frequencies this quantity increases as $f^{7/5}$. The flux keeps growing with the frequency since the amount of GWs for a fixed value of $h_c$ increases as $d\rho_{\rm gw}/df \propto f$ while the conversion probability decays slower.\\
In Fig.~\ref{fig: Angular Flux}, we show an illustrative example of the all sky flux map from a SGWB charactized by a characteristic strain $h_c=10^{-25}$ at frequency $f=10^{15}\, \rm Hz$, following the model with constant magnetic field. The model with decaying magnetic field predicts a $\mathcal{O}(0.001)$ weaker flux. Since the latter predicts a lower surface magnetic fields (see App.~\ref{App: Neutron Star Distribution}), less gravitons would be converted in the NS magnetospheres. For this reason, the flux from the decaying magnetic field model is typically smaller. 

\subsection{Upper Limits}\label{SubSec: upper limits}
The photon flux from this SGWB conversion must now be compared to observations of the all-sky integrated flux. In Fig.~\ref{fig: Flux vs observation} (right) we show the latter after the substraction of all galactic sources which are known and accounted for~\cite{Hill:2018trh}. Then, $F^{\rm galac.}_{\gamma}(f,h_c)$ can not be brighter than this flux, otherwise it would be in conflict with current observations. Thus, we can place an upper bound on the characteristic strain $h_c$ by demanding that our flux does not exceed the cosmic radiation background reported by various experiments. 
\begin{figure}[t!]
    \centering
    \includegraphics[width=0.9\textwidth]{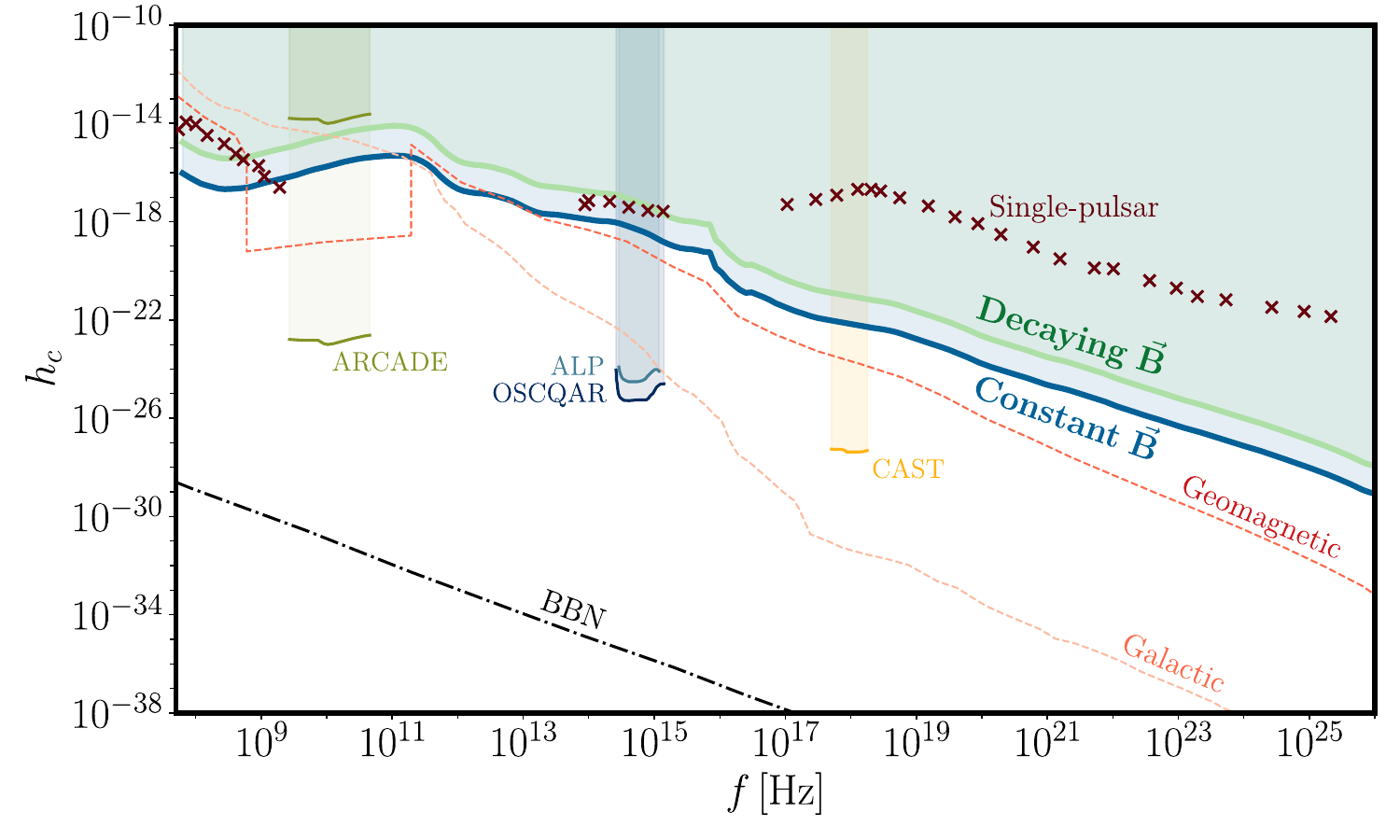}
    \caption{Constraints on the characteristic strain $h_c$ from different experiments. Blue and green solid lines represent our bounds from an induced photon flux in the magnetosphere of the galactic NSs, for NS models with a constant and decaying $\Vec{B}$, respectively. %The blue and green lines represent respectively the upper limits from the Model 1 and Model 2 of galactic NSs. The dashed blue and green lines show similar limits but assuming that all gravitons crossing the magnetosphere are converted with the same probability (see text). 
    We also show limits from OSQAR, ALP and CAST \cite{Ejlli:2019bqj} as well as the one from ARCADE (the upper and lower lines corresponding to uncertainty on the cosmic magnetic field) \cite{Domcke:2020yzq} and Big Bang Nucleosynthesis \cite{Cyburt:2015mya} (black dashed-dotted lines). The red and orange dashed curves represent the limits from conversion in the geomagnetic field and galactic magnetic field respectively \cite{Ito:2023nkq}. %Finally, we show how our limits improve the ones from a single pulsar analysis \cite{Ito:2023fcr} (red crosses) even when taking a conservative approach (solid lines). \TB{Which data points have you chosen for the single-pulsar analyses? There are a lot of them in the original figure, and no all of them seem to appear here.}
    Finally, the red crosses are the constraints from the Crab pulsar observation~\cite{Ito:2023fcr}. The code and data to reproduce this figure are publicly available \href{https://github.com/costa-francesco/High-frequency-GW-bounds}{\faGithub}.}
    \label{fig: Constraints}
\end{figure}
%

%From Eqs.\eqref{Eq: Single Pulsar Flux} and \eqref{Eq: Galactic Flux}, is it easy to see that the NSs induced flux $\partial F^{\rm galac.}_{\gamma}/\partial f$ scales as $\sim h_c^2$. Therefore, a too large characteristic strain would eventually be associated with a flux in conflict with current observations. This would naturally translate into upper limits on the characteristic strain $h_c$.\\
%As a first step into the derivation of those upper limits, we study on the left pannel of Fig.\ref{fig: Flux vs observation} the spectrum $f\times \partial F^{\rm galac.}_{\gamma}/\partial f$ normalized by $h_c^2$ so that it is no longer dependent on the characteristic strain. Interestingly, for large frequencies, the flux increases as a simple power law $\sim f ^{6/5}$. This is directly due to the fact that for a fixed characteristic strain $h_c$, the amount of gravitational wave scales as $ d\rho_{\rm gw}/df \sim h_c^2 f$, and therefore increases linearly with the frequency. Since the conversion probability at large frequencies only decreases as $\sim f^{-4/5}$ (see Eq.\eqref{Eq: Large freq}), the small chances to be converted are conter-balanced by a larger amount of gravitons.\\

%\noindent Intuitively, for each frequency, the maximal value for the characteristic strain $h_c$ is found by asking that this flux does not exceed the cosmic radiation background reported by various experiments. We show this background as a function of the frequency in the right pannel of Fig.\ref{fig: Flux vs observation} (see Ref.\cite{Hill:2018trh} for a summary of those).\\
The limits found with this method are reported as blue and green solid lines in Fig.~\ref{fig: Constraints}, for a constant and a decaying magnetic field, respectively. For comparison, we also show the current limits derived from axion experiments: OSQAR, ALP and CAST~\cite{Ejlli:2019bqj} as well as those derived using ARCADE~\cite{Domcke:2020yzq} (see Refs.\cite{Domcke:2023qle,Aggarwal:2020olq} for a summary of those constraints). The orange and red dashed curves show the upper limits derived when comparing the induced photon-flux produced in the geomagnetic field and galactic magnetic field~\cite{Ito:2023nkq} to the same telescope observations used in this work. Interestingly, we find our limits to be the best in the frequency range $(10^8-10^9)\, \mathrm{Hz}$ and comparable to the geomagnetic bounds around $(10^{12}-10^{15})\, \mathrm{Hz}$. In contrast with bounds from geomagnetic conversion, our results predict a flux which peaks in the Galactic Center, as shown in Fig.~\ref{fig: Angular Flux}. Observations of the galactic flux angular distribution or focused on the Galactic Center could improve our bounds with respect to others. At larger frequencies, the induced photon flux from galactic NSs is systematically smaller than the one from the geomagnetic field and galactic magnetic field. For this reason our limits are less competitive at large frequencies. \\ 
Comparable calculations to the ones presented in this work have been also performed in the case of the Crab pulsar, with data obtained from single-pulsar observations. In Ref.~\cite{Ito:2023fcr}, the induced photon-flux has been compared to the current observations of this NS and led to the red crosses in Fig.~\ref{fig: Constraints}. At least for present data, using the galactic population of NSs at large frequencies increases by several orders of magnitudes the constraints compared to a single pulsar analysis.\\
Finally, the black dashed-dotted line shows the upper limit on the amount of gravitational waves produced before Big Bang Nucleosynthesis (BBN), which would contribute to the number of relativistic degrees of freedom~\cite{Cyburt:2015mya}. Since this limit does not apply to post-BBN or monochromatic sources, current bounds are important even if they fall short compared to BBN constraints. Still, these bounds correspond to GWs with $d\Omega_{\mathrm{gw}}/d\log f\gg 1$. Such sources would need to be extremely monochromatic in order not to overclose the Universe. Even if they emitted in a more reasonable frequency width, no known cosmological sources can produce HFGWs intense enough to saturate our bounds. The mergers of exotic compact objects or cosmological sources could produce GWs with the right frequency, but with $\Omega_{\mathrm{gw}}<1$~\cite{Giudice_2016,Domcke:2020yzq,Maggiore:2007book,Franciolini:2022htd,Aggarwal:2020olq}.
Therefore, further improvement is still required for all state-of-the-art HFGWs analysis to constrain such sources. \\ 
With respect to our results, the bounds here presented will be improved by more precise all-sky observations. It might also be possible to profit from the accumulation of neutron stars in the Galactic center direction. However, such backgrounds are harder to model and a lower signal-to-noise ratio would make an improvement challenging. \\
Again, these kind of analysis would only constrain homogeneous fluxes, either of cosmological or astrophysical (but distributed around all the sky) origin. Single localized events would require a perfect alignment between the NS and the GW source and require a different approach to an all-sky observation. Thus, these are beyond the scope of our work.

\section{Conclusion}\label{Sec: Conclusion}

In this paper we presented upper limits on the high frequency SGWB from the (inverse) Gertsenshtein effect in the magnetosphere of the galactic NSs. This is particularly important since this region of the GW spectrum is only partially constrained and may be used to probe new physics phenomena.\\
\indent As a first step, we recalled the graviton-photon conversion in the NS magnetosphere. An effective conversion can only happen in regions where the effective photon mass is low enough so that both energy and momentum are conserved in the process. Due to the Cotton-Mouton effect, the photon polarization perpendicular to the external magnetic field is less massive compared to the parallel one which translates into a significantly larger conversion of the $+$ polarized gravitons compared to the $\times$ ones (as also observed by~\cite{Ito:2023fcr}). For both polarizations, we derived analytical formulae for the conversion probability. From this, we extracted the induced photon flux from a single NS and then from the ensemble of galactic NSs using two realistic population models~\cite{Faucher-Giguere:2005dxp,Popov:2009jn}. The first assumes a decaying magnetic field whereas the second considers a constant surface magnetic field over time. Since the graviton-photon conversion is enhanced by large magnetic fields, we find that the second model predicts a larger induced photon flux. Finally, we compared this induced photon flux to several telescope observations. Asking that the flux does not exceed the current observations provides upper limits on the SGWB.\\
\indent Our conclusions are the following: first, since many phenomena could induce a SGWB, it is highly expected for the galactic NSs to be an important site of graviton-photon conversion. However, only a large-enough high frequency SGWB would induce a photon signal which can be potentially detectable with future experiments. Such SGWB could be explained by mergers of light primordial black holes. Secondly, we derived competitive upper limits in the frequency range $\sim (10^8 $ - $10^{15}) \, \rm Hz$, confirming, and improving in some frequency range, the existing constraints from similar studies on the geomagnetic and galactic magnetic field~\cite{Domcke:2020yzq,Ito:2023nkq}. Beyond $\sim 10^{15} \, \rm Hz$, we found the induced photon flux from galactic NSs to be significantly smaller than the one from the galactic magnetic field~\cite{Ito:2023nkq}. This naturally translates into weaker constraints.

\section*{Acknowledgements}
We thank Valerie Domcke, Jordi Salvadó and Jinsu Kim for useful comments on the manuscript and Nicolas Grimbaum, Asuka Ito, Kazunori Kohri, Kazunori Nakayama for useful discussion.
TB acknowledges financial support from the Spanish grants PID2022-136224NB-C21, PID2019-108122GBC32, PID2019-105614GB-C21, and from the State Agency for Research of the Spanish Ministry of Science and Innovation through the ``Unit of Excellence María de Maeztu 2020-2023'' award to the Institute of Cosmos Sciences (CEX2019-000918-M). The work of F.C. is supported by the European Union’s Horizon 2020 research and innovation programme under the Marie Sklodowska-Curie grant agreement No 860881-HIDDeN.

\appendix

\section{Neutron Star Distribution}\label{App: Neutron Star Distribution}

To accurately predict the total induced photon flux on Earth, a realistic distribution for the properties of the NSs must be used (surface magnetic field $B$ \footnote{Later on, $B$ will always refer to the surface magnetic field.} and spin period $T$).\\
Numerous NS observations have been performed (more than 2000 rotation-powered NSs \cite{Manchester:2001fp,Edwards:2001de}). Based on those, recent efforts have been made in developing NS models (initial magnetic field and period distributions as well as their time evolution) capable of reproducing the observed data. In this appendix, we summarized the ideas of two distinct models that are used in the main text \cite{Faucher-Giguere:2005dxp,Popov:2009jn}.\\
In the first (Model 1, referred to in the main text as "constant magnetic field"), the NS magnetic field is assumed constant and therefore fixed to its initial birth value. On the other hand, the spin period increases in time as the NS loses energy by dipole radiation and plasma effects (see Refs.\cite{Spitkovsky:2006np,Philippov:2013aha,Johnston:2017wgm}). It is found that the period evolves as \cite{Safdi:2018oeu}
\begin{equation}
    T(t) = T_0\sqrt{1+\frac{4t}{3\tau_0}},
\end{equation}
with
\begin{equation}
    \tau_0 \sim 10^4\left(\frac{10^{12} \rm G}{B}\right)^2\left(\frac{T_0 }{0.01 \rm s}\right)^2 \rm yr,
\end{equation}
where $T_0$ is the initial spin period and $B$ the constant magnetic field of the NS.\\
Eventually, as the period increases, the NS spin might be low enough to prevent a coherent radio-pulsed emission. At this stage, the NS becomes undetectable (further on referred to as "dead" or "inactive" NSs, although it is important to stress that it does not imply that it loses its magnetic field) \cite{Zhang:2000rd}. This transition is expected to happen when the ratio $B/T^2$ is smaller than \cite{Bhattacharya:1992vf}
\begin{equation}
    B/T^2 < 0.34 \times 10^{12}\rm G\, s^{-2}. 
\end{equation}
Post-death, NSs are expected to lose energy only through dipole radiation \cite{Safdi:2018oeu}, this translates into a modified period evolution,
\begin{equation}
    T(t>t_{\rm death}) = T(t_{\rm death})\sqrt{1+\tan^2 \alpha \left(1-e^{-\frac{4(t-t_{\rm death})\cos^2\alpha}{3\tau_0}}\right)},
\end{equation}
where $t_{\rm death}$ is the time at which the NS becomes inactive and $\alpha$ is the misalignment angle between the NS rotation and magnetic axis (see ref.\cite{Safdi:2018oeu}).\\
\begin{table}[t!]
\centering
%\begin{tabular}{ |p{3cm}\|p{3cm}|p{3cm}|p{3cm}|  }
\begin{tabular}{ |p{3cm}|p{4cm}|p{4cm}| }
 \hline
 \multicolumn{3}{|c|}{\textbf{Initial Distributions}} \\ [1em]
 \hline
 \hline
  \textbf{Model} &  \textbf{Variance [G], [s]} & \textbf{Mean [G], [s]}\\ [1em]
 \hline
 \hline
 Model 1 & $\sigma_B = 0.55, \, \sigma_T = 0.15$ & $\mu_B=12.95, \, \mu_T =0.3$ \\ [0.5em]
 \hline 
 Model 2 & $\sigma_B = 0.6, \, \sigma_T = 0.1$ & $\mu_B=13.25, \, \mu_T =0.25$ \\[0.5em]
 \hline
\end{tabular}
\caption{Variance and mean values of the log-normal distributions \eqref{Eq: Log-normal} for the initial surface magnetic field $B_0$ and spin period $T_0$ as derived in Refs.\cite{Faucher-Giguere:2005dxp,Popov:2009jn}.}
\label{Tab: Initial Distribution}
\end{table}
In the second model (Model 2, referred in the main text as "decaying magnetic field"), the spin period time evolution is considered to be the same as in Model 1. Nonetheless, the magnetic field is here considered decaying in time. This is the essential difference between these two models. For the magnetic field decay, we follow Ref.\cite{Safdi:2018oeu} and describe the time evolution of the surface magnetic field as 
\begin{equation}
    \frac{dB}{dt} = -B \left( \frac{1}{\tau_{\rm Ohm}} + \left(\frac{B}{B_0}\right)^2\frac{1}{\tau_{\rm Ambip}}\right),
\end{equation}
where $B_0$ is the initial magnetic field and $\tau_{\rm Ambip}$, $\tau_{\rm Ohm}$ are the decay rate of the magnetic field by ambipolar effects \cite{1992ApJ...395..250G,1995MNRAS.273..643S} and ohmic dissipation \cite{1990A&A...229..133H}, respectively.\\

\begin{figure}[t!]
    \centering
    \includegraphics[width=1\textwidth]{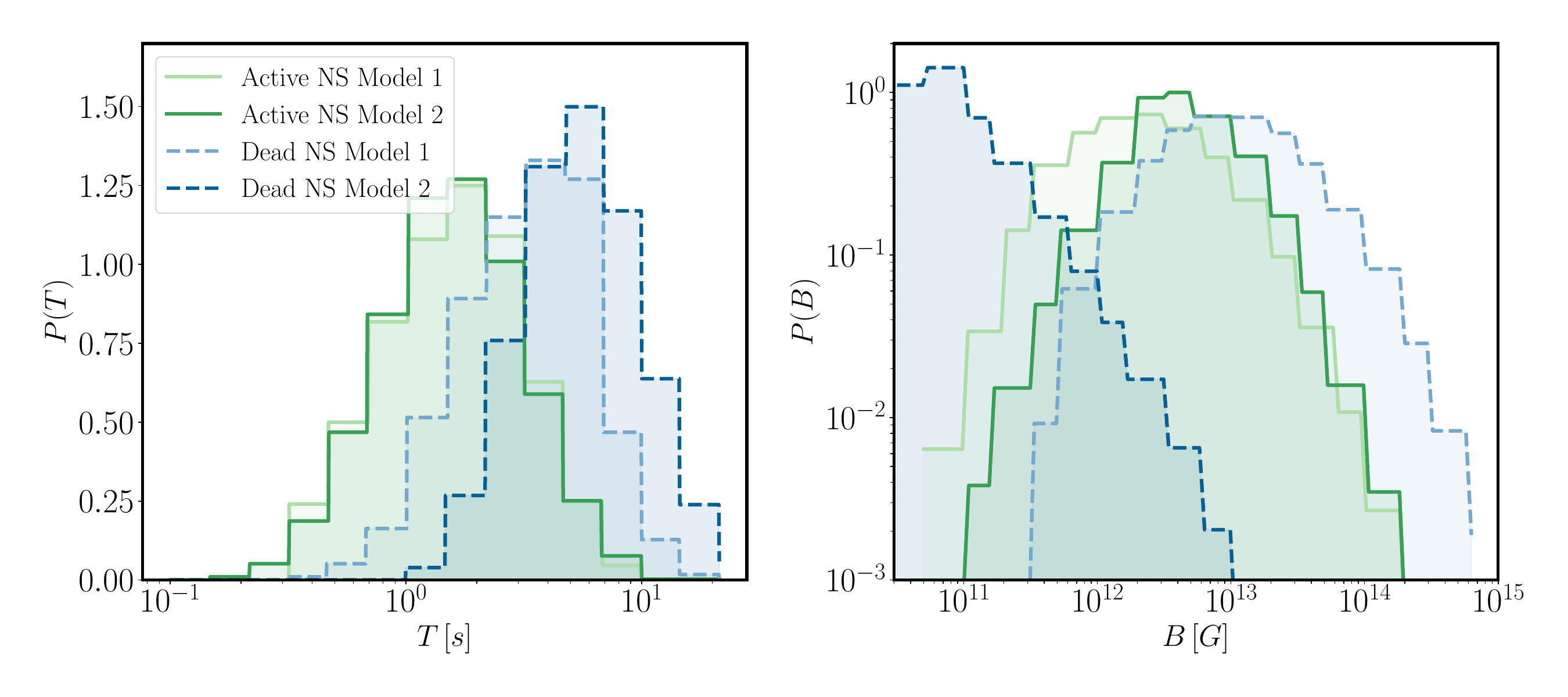}
    \caption{\textit{Left}: Today's distribution for the spin period $T$ of the active (green) and dead (blue) NSs. Model 1 and Model 2 are represented by the light and dark colors respectively. \textit{Right}: Same for the surface magnetic field $B$. In both cases, the distributions are normalized so that $\int d \log(x) P(x) =1 $. This figure is adapted from Fig.4 of Ref.\cite{Safdi:2018oeu}.}
    \label{fig: Distribution NSs}
\end{figure}

In Refs.\cite{Safdi:2018oeu,Faucher-Giguere:2005dxp,Popov:2009jn}, a large population of NSs have been simulated and evolved until today using the two above models. In order to reproduce the observational data \cite{Manchester:2001fp,Edwards:2001de}, the initial distributions for the surface magnetic field $B_0$ and period $T_0$ have been found to be given by,
\begin{equation}\label{Eq: Log-normal}
\begin{split}
    P(B_0) &= \frac{1}{\sqrt{2\pi}\,B_0\sigma_{B}} \exp\left[-\frac{\log^2(B_0/\mu_B)}{(2\sigma_B^2)}\right],\\
    P(T_0) &= \frac{1}{\sqrt{2\pi}\,T_0\sigma_{T}} \exp\left[-\frac{\log^2(T_0/\mu_T)}{(2\sigma_T^2)}\right],
\end{split}
\end{equation}
where $(\sigma_B,\mu_B)$ and $(\sigma_T,\mu_T)$ are given for Model 1 and Model 2 in Tab.\ref{Tab: Initial Distribution}.\\
\noindent Furthermore, from the same simulations, the today's magnetic field and period distributions can be extracted. In both models, it has been found that only $\sim 0.4\%$ of the NSs are still active \cite{Safdi:2018oeu}. In Fig.\ref{fig: Distribution NSs}, we show today's distributions for both the surface magnetic field and the spin period in the two models. In particular, we show separately the distributions (normalized) for the active and dead NSs. It can be appreciated that, in both models, the dead NSs have larger periods. This is directly related to the fact that dead NSs have more time to evolve their spin period to larger values. On the right panel, the magnetic field distributions for the active and dead NSs in Model 1 are relatively similar with slightly larger values for the dead NSs. This is due to the fact that for larger magnetic fields, $\tau_0$ gets smaller making the spin period decrease faster. Interestingly, the magnetic field distributions for the Model 2 are quite different. Recalling that in this model, the magnetic field is assumed to decay in time, the dead NSs, assumed to be older than the active ones, have more time to decay and for this reason, their surface magnetic field is significantly reduced compared to the active population (assumed younger).\\
There exist different methods to determine the number of neutron stars in the Milky Way. On the one hand, supernovae explosions, which lead into to NS, release an important amount of iron. The study of the chemical composition of the galaxy in the Milky Way and its iron content suggests that about $10^9$ NSs have been produced~\cite{Arnett1989}. 
On the other hand, the disk stellar birth rate has been observed to be approximately constant over the galactic history~\cite{Rocha-Pinto2000}. This implies a roughly constant NS formation rate as well, which leads to $4 \times 10^8 $ NSs in the disk. The stellar population in the bulge is expected to have formed in a much shorter time scale around 12 Gyrs ago, and its stellar mass function is skewed toward high masses~\cite{Ballero2007}. The early time population of heavy stars led to a more important NS population in the bulge,  around $6 \times 10^8$~\cite{Ofek:2009wt}. All in all, this leads again to the estimation of $10^9$ NSs in the galaxy.
%The total number of NSs can be estimated from the chemical composition of the Galaxy. In particular, its iron content provides the information that about $10^9$ NS have been produced \cite{Ofek:2009wt}. Similar estimation can be found from the rate of core-collapse supernovae observed in our current days and assuming that this rate has been constant over the lifetime of the galaxy \cite{Sartore:2009wn}. \\
We follow standard choices in the literature and take $N_{\rm bulge}  = 4.8\times 10^8$ and $N_{\rm disk} = 3.2\times 10^8$ NSs \cite{Ofek:2009wt,Sartore:2009wn}. Of course, depending of the total number of NSs, the upper limits derived in this work would change. It is expected that increasing (decreasing) the number of NSs would translate into stronger (weaker) constraints. \\
Each population follows a different spatial distribution. The disk distribution (in a frame with the origin at the galactic center) has been extracted from millisecond pulsar observations \cite{Lorimer:2006qs,Bartels:2018xom} (see also \cite{Faucher-Giguere:2009fex} for another parametrisation), 
\begin{equation}\label{Eq: Disk}
    n_{\rm disk} (r,z) = N_{\rm disk}\frac{C^{B+2}e^{-C}}{4\pi r_{\odot}^2\sigma_z\Gamma(B+2)}\left(\frac{r}{r_{\odot}}\right)^B e^{-C\frac{r-r_{\odot}}{r_{\odot}}}e^{-\mid z \mid/\sigma_z},
\end{equation}
where $B=3.91$, $C=7.54$, $\sigma_z=0.76 \rm kpc$ and $r_{\odot}$ is the sun distance from the galactic center. The bulge distribution is taken from Ref.~\cite{Edwards:2020afl} and is assumed to follow the stellar distribution \cite{Binney:1996sv,Bissantz:2001wx},
\begin{equation}\label{Eq: Bulge}
    n_{\rm bulge} (r,z) = N_{\rm bulge}\frac{11.1}{\rm kpc^3}\frac{e^{-(r'/r_{\rm cut})^2}}{(1+r'/r_0)^{\lambda}},
\end{equation}
with $r' = \sqrt{r^2+(z/0.5)^2}$, $r_0 = 0.075\,  \rm kpc $, $\lambda = 1.8$ and $r_{\rm cut} = 2.1\,  \rm kpc$.\\
As already mentioned, from this spatial distribution, $0.4\%$ will be active NSs with surface magnetic field and spin period distributed as in Fig.\ref{fig: Distribution NSs} (green solid lines) and $99.6\%$ will be inactive NSs with surface magnetic field and spin period distributed as the blue dashed lines in Fig.\ref{fig: Distribution NSs}.  In the main text, the total flux on Earth from graviton-to-photon conversion in the NS magnetospheres has been calculated for both Model 1 and Model 2. For both models, we assumed an NS radius of $r_0 = 10$ km REF.

\bibliographystyle{JHEP_improved}
\bibliography{./Biblio}

\end{document}